\documentstyle[12pt]{article}
\input epsf.sty
\catcode`@=11
\def\chkspace{%
  \relax   % Calm down any expanding \if's.
  \begingroup\ifhmode\aftergroup\dochksp@ce\fi\endgroup}
\def\dochksp@ce{%
  \unskip              % Remove any preceding horizontal glue
  \futurelet\chkspct@k\d@chkspc  % Grab the next token and look ahead
}
\def\d@chkspc{%
  \let\nxtsp@ce=\relax
  \ifx\chkspct@k.\else     % Don't put spaces before punctuation ...
    \ifx\chkspct@k,\else
      \ifx\chkspct@k;\else
        \ifx\chkspct@k!\else
          \ifx\chkspct@k?\else
            \ifx\chkspct@k:\else
              \ifx\chkspct@k)\else
              \ifx\chkspct@k(\else
                \ifx\chkspct@k]\else
                  \ifx\chkspct@k-\else
                    \ifx\chkspct@k\egroup\else  % or close groups.
                      \let\nxtsp@ce=\put@space  % Put a space everywhere else.
                    \fi
                  \fi
                \fi
              \fi
              \fi
            \fi
          \fi
        \fi
      \fi
    \fi
  \fi
  \nxtsp@ce
}
\def\put@space{$\;$}
\catcode`@=12

\def\ra{\relax\ifmmode \rightarrow\else{{$\rightarrow$}}\fi\chkspace}
\def\Ra{\relax\ifmmode \Rightarrow\else{{$\Rightarrow$}}\fi\chkspace}
\def\etal{{\it et al.}\chkspace}

\def\ie{{\it i.e.}\chkspace}

\def\ep{{e$^+$e$^-$}\chkspace}

\def\qu{\relax\ifmmode \quad\else{{$\quad$}}\fi\chkspace}

\def\gluino{\relax\ifmmode \tilde{g} \else $\tilde{g}$ \fi\chkspace}

\def\qq{\relax\ifmmode q\overline{q}
\else $q\overline{q}$ \fi\chkspace}
\def\ff{\relax\ifmmode f\overline{f}
\else $f\overline{f}$ \fi\chkspace}

\def\bbar{$\overline{b}$\chkspace}

\def\bb{\relax\ifmmode b\bar{b}
       \else $b\bar{b}$ \fi\chkspace}
\def\ccrm{\relax\ifmmode {\rm c}\bar{\rm c}
       \else ${\rm c}\bar{\rm c}$ \fi\chkspace}

\def\tt{\relax\ifmmode {\rm t}\bar{\rm t}
       \else ${\rm t}\bar{\rm t}$ \fi\chkspace}
\def\ss{\relax\ifmmode {\rm s}\bar{\rm s}
       \else ${\rm s}\bar{\rm s}$ \fi\chkspace}
\def\uu{\relax\ifmmode {\rm u}\bar{\rm u}
       \else ${\rm u}\bar{\rm u}$ \fi\chkspace}
\def\dd{\relax\ifmmode {\rm d}\bar{\rm d}
       \else ${\rm d}\bar{\rm d}$ \fi\chkspace}

\def\qqg{\relax\ifmmode q\overline{q}g
\else $q\overline{q}g$ \fi\chkspace}
\def\bbg{\relax\ifmmode b\overline{b}g
\else $b\overline{b}g$ \fi\chkspace}
\def\ccg{\relax\ifmmode c\overline{c}g
\else $c\overline{c}g$ \fi\chkspace}
\def\ttg{\relax\ifmmode t\overline{t}g
\else $t\overline{t}g$ \fi\chkspace}

\def\afb{\relax\ifmmode A_{FB} \else
{{$A_{FB}$}}\fi\chkspace}
\def\afbb{\relax\ifmmode A_{FB}^b \else
{{$A_{FB}^b$}}\fi\chkspace}
\def\pafb{\relax\ifmmode \tilde{A}_{FB} \else
{{$\tilde{A}_{FB}$}}\fi\chkspace}
\def\pafbb{\relax\ifmmode \tilde{A}_{FB}^b \else
{{$\tilde{A}_{FB}^b$}}\fi\chkspace}

\def\pafbzo{\relax\ifmmode \tilde{A}_{FB}|_{O(0)} \else
{{$\tilde{A}_{FB}|_{O(0)}$}}\fi\chkspace}
\def\pafbfo{\relax\ifmmode \tilde{A}_{FB}|_{\oalp} \else
{{$\tilde{A}_{FB}|_{\oalp}$}}\fi\chkspace}
\def\pafbso{\relax\ifmmode \tilde{A}_{FB}|_{\oalpsq} \else
{{$\tilde{A}_{FB}|_{\oalpsq}$}}\fi\chkspace}
\def\pafbto{\relax\ifmmode \tilde{A}_{FB}|_{\oalpc} \else
{{$\tilde{A}_{FB}|_{\oalpc}$}}\fi\chkspace}

\def\pafbbzo{\relax\ifmmode \tilde{A}_{FB}^b|_{O(0)} \else
{{$\tilde{A}_{FB}^b|_{O(0)}$}}\fi\chkspace}
\def\pafbbfo{\relax\ifmmode \tilde{A}_{FB}^b|_{\oalp} \else
{{$\tilde{A}_{FB}^b|_{\oalp}$}}\fi\chkspace}
\def\pafbbso{\relax\ifmmode \tilde{A}_{FB}^b|_{\oalpsq} \else
{{$\tilde{A}_{FB}^b|_{\oalpsq}$}}\fi\chkspace}
\def\pafbbto{\relax\ifmmode \tilde{A}_{FB}^b|_{\oalpc} \else
{{$\tilde{A}_{FB}^b|_{\oalpc}$}}\fi\chkspace}

\def\afbo0{\tilde{A}_{FB}|_{O(0)}}
\def\afbo1{\tilde{A}_{FB}|_{\oalp}}
\def\afbo2{\tilde{A}_{FB}|_{\oalpsq}}
\def\afbo3{\tilde{A}_{FB}|_{\oalpc}}

\def\lam{\relax\ifmmode \Lambda_{\overline{MS}}
       \else {{$\Lambda_{\overline{MS}}$}}\fi\chkspace}
\def\lamuds{\relax\ifmmode \Lambda^{(3)}_{\overline{MS}}
       \else {{$\Lambda^{(3)}_{\overline{MS}}$}}\fi\chkspace}
\def\lamudsc{\relax\ifmmode \Lambda^{(4)}_{\overline{MS}}
       \else $\Lambda^{(4)}_{\overline{MS}}$\fi\chkspace}
\def\lamudscb{\relax\ifmmode \Lambda^{(5)}_{\overline{MS}}
       \else $\Lambda^{(5)}_{\overline{MS}}$\fi\chkspace}

\def\alp{\relax\ifmmode \alpha_s\else $\alpha_s$\fi\chkspace}
\def\alpbar{\relax\ifmmode \bar{\alpha_s}
       \else $\bar{\alpha_s}$\fi\chkspace}
\def\alpmz{\relax\ifmmode \alpha_s(M_Z)\else $\alpha_s(M_Z)$\fi\chkspace}
\def\alpmzsq{\relax\ifmmode \alpha_s(M_Z^2)
       \else $\alpha_s(M_Z^2)$\fi\chkspace}

\def\oalp{\relax\ifmmode O(\alpha_s)\else{{O($\alpha_s$)}}\fi\chkspace}
\def\oalpsq{\relax\ifmmode O(\alpha_s^2)
           \else{{O($\alpha_s^2$)}}\fi\chkspace}
\def\oalpc{\relax\ifmmode O(\alpha_s^3)
           \else{{O($\alpha_s^3$)}}\fi\chkspace}
\def\oalpf{\relax\ifmmode O(\alpha_s^4)
           \else{{O($\alpha_s^4$)}}\fi\chkspace}

\def\rb{\relax\ifmmode R_3^b/R_3^{all}
           \else{{$R_3^b/R_3^{all}$}}\fi\chkspace}
\def\rc{\relax\ifmmode R_3^c/R_3^{all}
           \else{{$R_3^c/R_3^{all}$}}\fi\chkspace}
\def\ruds{\relax\ifmmode R_3^{uds}/R_3^{all}
           \else{{$R_3^{uds}/R_3^{all}$}}\fi\chkspace}
\def\ri{\relax\ifmmode R_3^i/R_3^{all}
           \else{{$R_3^i/R_3^{all}$}}\fi\chkspace}
\def\rj{\relax\ifmmode R_3^j/R_3^{all}
           \else{{$R_3^j/R_3^{all}$}}\fi\chkspace}
\def\alpi{\relax\ifmmode \alpha^i_s/\alpha^{all}_s
           \else{{$\alpha^i_s/\alpha^{all}_s$}}\fi\chkspace}

\def\mbz{\relax\ifmmode m_b(M_Z)
           \else{{$m_b(M_Z)$}}\fi\chkspace}
\def\mbb{\relax\ifmmode m_b(M_b)
           \else{{$m_b(M_b)$}}\fi\chkspace}

\def\prd{Phys. Rev.\chkspace}

\def\z0{\relax\ifmmode Z^0 \else {$Z^0$} \fi\chkspace}
\def\h0{\relax\ifmmode H^0 \else {$H^0$} \fi\chkspace}
\def\Dst{\relax\ifmmode {\rm D}^* \else {D$^*$}\fi\chkspace}
\def\Dpl{\relax\ifmmode {\rm D}^+ \else {D$^+$}\fi\chkspace}
\def\D0{\relax\ifmmode {\rm D}^0 \else {D$^0$}\fi\chkspace}
\def\Kst{\relax\ifmmode {\rm K}^* \else {K$^*$}\fi\chkspace}
\def\K0{\relax\ifmmode {\rm K}^0_s \else {K$^0_s$}\fi\chkspace}
\def\Kpl{\relax\ifmmode {\rm K}^+ \else {K$^+$}\fi\chkspace}
\def\Kstz{\relax\ifmmode {\rm K}^{*0} \else {K$^{*0}$}\fi\chkspace}

\def\beq{\begin{equation}}
\def\eeq{\end{equation}}
\def\bea{\begin{eqnarray}}
\def\eea{\end{eqnarray}}

\begin{document}

{\hfill{SLAC-PUB-8302}}

{\hfill{OUNP-99-14}}

{\hfill{October 1999}}

\vskip 1.5truecm

\begin{center}

\Large\bf

{Searches for Anomalous Effects in the \bbg Coupling$^{**}$}

\vskip 1truecm

{\large

P N Burrows$^{\dag}$
             
\vskip .5truecm

{\it Representing the SLD Collaboration$^*$}

{Stanford Linear Accelerator Center,}

{Stanford University, Stanford, CA 94309}

}

\vskip 1truecm

\end{center}

\centerline{\bf Abstract}

\noindent
The unique SLD CCD vertex detector combined with the 
highly-polarised electron beam 
allow us to search for an anomalous chromomagnetic
coupling of the $b$-quark, as well as P-odd, T$_N$-odd and CP-odd processes at 
the \bbg vertex. 

\vfill
\noindent
{\it Talk presented at the International Europhysics Conference on High Energy Physics,
Tampere, Finland, 15-21 July 1999.}

\vskip .5truecm

\noindent
{$^{\dag}$ 
Particle \& Nuclear Physics, Keble Rd., Oxford, OX1 3RH, UK;
E-mail: {p.burrows@physics.ox.ac.uk}.
Supported by the UK Particle Physics \& Astronomy Research Council
}

\vskip .5truecm

\noindent
{$^{**}$ Work supported by Department of Energy contract DE-AC03-76SF00515 (SLAC).}
\eject

\section{Introduction}

One expects new high-mass-scale 
dynamics to couple preferentially to the massive third-generation fermions.
The study of \ep\ra\bb events is hence of considerable interest.
Using inclusive \z0\ra\bb decays one can measure 
$R_b$ = $\Gamma(Z^0\rightarrow$\bb)/$\Gamma(Z^0\rightarrow$\qq) 
and the \z0-$b$ parity-violating coupling $A_b$.
In recent years several reported determinations of
these quantities have differed from Standard Model (SM) expectations 
at the few $\sigma$ level, 
arousing considerable interest and speculation. 
Currently $R_b$ is in good agreement with the SM, whereas
$A_b$ appears to be about 2.5$\sigma$ low~\cite{mnich}.

We have therefore investigated in detail the
strong-interaction dynamics of the $b$-quark.
We have compared the strong coupling of the gluon to
$b$-quarks with that to light- and charm-quarks~\cite{sldflav} and
found all couplings to be equal to within the experimental sensitivity
of a few per cent.
We have also studied the structure of 3-jet \bbg events~\cite{sldbbgpub},
as well as
tested parity (P) and charge$\oplus$parity (CP) conservation at the \bbg 
vertex;
here we present a preliminary update of the latter two measurements using a data
sample more than 3 times larger.
Full details can be found in~\cite{sldbbgnew, sldsymm}.

We used
roughly 550,000 hadronic \z0 decays recorded between 1993 and 1998 
in the SLC Large Detector (SLD). The average magnitude of the electron-beam 
polarisation was 73\%.
We applied iterative clustering algorithms to select 3-jet events.
In order to improve the energy resolution 
the jet energies were rescaled kinematically according to the angles
between the jet axes, assuming energy and momentum conservation and massless
kinematics. 
The jets were then labelled in order of energy such that $E_1 > E_2 > E_3$.

\section{\bbg Observables and Tagging Strategy}

We considered the following \bbg observables: 

\noindent
$\bullet$
the scaled gluon energy, $x_g=E_{gluon}/E_{beam}$,
to test for anomalous \bbg 
couplings, and the gluon polar angle w.r.t. the e$^-$ beam, $\theta_g$; 

\noindent
$\bullet$
the $b$-quark polar angle, $\theta_b$, and azimuthal angle, $\chi$ (between the
$b$-quark-gluon plane and the $b$-quark-e$^-$ plane),
to test for parity-violation at the \bbg vertex; 

\noindent
$\bullet$
the polar angle, $\omega$, of the normal to the \bbg plane to test
for T$_N$-odd final-state interactions.
With the normal defined by $\vec{p_{b_i}}\times \vec{p_{b_j}}$ ($|p_{b_i}|>|p_{b_j}|$) the
forward-backward asymmetry of the angular distribution (see section 4) is CP-even;
with the normal defined by $\vec{p_b}\times \vec{p_{\bar{b}}}$ it is CP-odd.

In order to define these observables we require two different tagging strategies:
1) which jet is the gluon? \ie we need to tag two jets as being $b$ or \bbar; 
2) which jet is the gluon, and which is the $b$-quark? 
\ie we need in addition to distinguish between the $b$ and \bbar jets.

In strategy 1 we reconstructed jets using the JADE algorithm with a
scaled-invariant-mass criterion $y_{cut}$ = 0.02; 
126,871 3-jet events were selected. 
Charged tracks with a large transverse impact parameter ($d$) w.r.t. the
interaction point were used to tag \bbg events~\cite{sldbbgnew}. 
The flavour tag was based on the number of `significant' 
tracks per jet, $N_{sig}^{jet}$, with $d/\sigma_d\geq3$. 
8196 events were selected in which exactly two jets were $b$-tagged 
by requiring each to have $N_{sig}^{jet}\geq2$ and in which
the remaining jet had $N_{sig}^{jet}<2$ and was hence tagged as the gluon. 
The efficiency for selecting true \bbg events is 12\%.
The inclusive gluon purity of the tagged-jet sample is 93\%. 

In strategy 2 we reconstructed jets using the Durham algorithm and $y_{cut}$ = 0.005; 
roughly 75,000 3-jet events were selected. 
A topological algorithm was applied to the set of
tracks in each jet to search for a secondary decay vertex. 
An event was selected as \bbg if at least one jet contained a vertex with 
invariant mass larger than 1.5 GeV/$c^2$~\cite{sldsymm}. 
14,658 events satisfied 
this requirement. With the new SLD VXD this selection is 84\% efficient for 
identifying a sample of \bbg events with 84\% purity, and containing 14\%
\ccg and 2\% light-flavor backgrounds.     
Jet 1 was chosen as the gluon jet only if it contained no significant track
and both jets 2 and 3 contained at least one such track. Jet 2 was chosen as the
gluon jet if it contained no significant track and jet 3 contained 
at least one significant track.  
Otherwise jet 3 was tagged as the gluon.
The momentum-weighted  charge was calculated for each `$b$' jet;
if the difference in charge, 
$Q_i - Q_j$, was negative (positive) jet $i$ was tagged as the $b$-jet
(\bbar-jet). 
The probability of correctly identifying the $b$-jet charge is 68\%.    

\section{Anomalous \bbg Chromomagnetic Coupling}
                                
We formed the distributions of $x_g$ and $\theta_g$. 
The non-\bbg-event backgrounds were subtracted, and the 
distributions were corrected for the
efficiency for accepting true \bbg events into the tagged sample, as well as
for bin-to-bin migrations caused by hadronisation, 
the resolution of the detector, and bias of the jet-tagging technique.
Fig.~1 shows the fully-corrected normalised distributions. 

\begin{figure}[h]
\vspace*{-0.3cm}
\begin{center}
\epsfxsize=12cm
\epsfysize=8cm
\epsffile{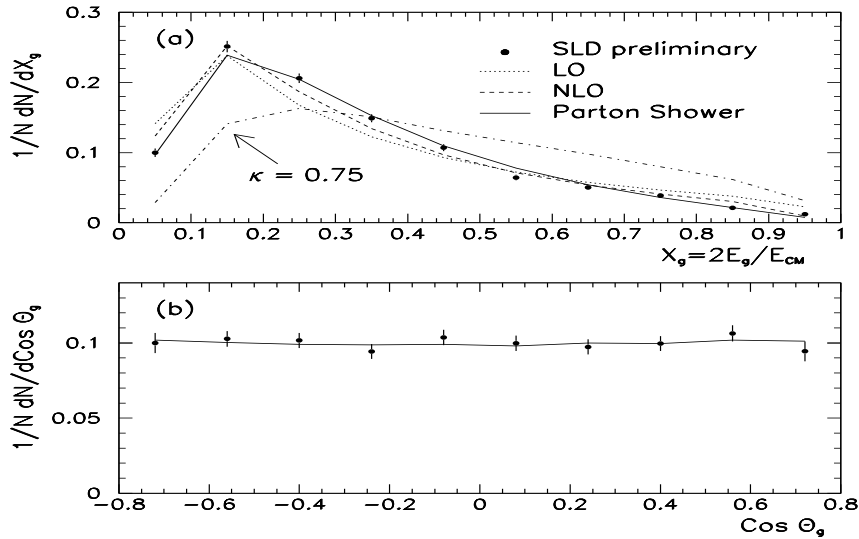}
\vspace*{-0.7cm}
\caption{(a) $x_g$, (b) cos$\theta_g$ (dots);
errors are statistical.
PQCD predictions (see text) are also shown.
}
\end{center}
\label{fig:bbg}
\vspace*{-0.9cm}
\end{figure}

We compared the data with PQCD predictions for
the same jet algorithm and $y_c$ value. We used leading-order (LO) and
NLO results based on recent calculations~\cite{arnd} 
in which quark mass effects were explicitly taken into
account; a $b$-mass value of $m_b(m_Z)=3$GeV/$c^2$ was used.
We also derived these distributions using the `parton 
shower' (PS) implemented in JETSET 7.4. 
The calculations reproduce the measured cos$\theta_g$ distribution,
which is insensitive to the details of higher-order soft parton
emission. For $x_g$, although the LO 
calculation reproduces the main features of the shape of the distribution,
it yields too few events in the region $0.2<x_g<0.5$, and too many events
for $x_g<0.1$ and $x_g>0.5$. The NLO calculation is noticeably better, but
also shows a deficit for $0.2<x_g<0.4$. The PS calculation describes the
data across the full $x_g$ range.
This suggests that multiple
orders of parton radiation need to be included.
We conclude that PQCD in the PS approximation accurately reproduces 
the gluon distributions in \bbg events. 

In QCD the quark chromomagnetic moment is induced at the 
one-loop level and is of order $\alpha_s$/$\pi$. 
A more general $\bbg$ Lagrangian term with a modified coupling
may be written:  
$$
{\cal L}^{b\overline{b}g} =  g_s\overline{b}T_a \{ \gamma_{\mu} + 
\frac{i\sigma_{\mu\nu}k^{\nu}}{2m_b}(\kappa - i \tilde{\kappa}\gamma_5)\} 
bG_a^{\mu}
$$
\noindent
where 
$\kappa$ and $\tilde{\kappa}$ parameterize the anomalous chromomagnetic and chromoelectric
moments, respectively, which might arise from physics beyond the SM. 
The effects of $\tilde{\kappa}$ are sub-leading w.r.t. those of $\kappa$, 
so for convenience we set $\tilde{\kappa}$ to zero. A non-zero $\kappa$
would modify the $x_g$ distribution in $\bbg$ events relative to the
standard QCD case. 
In each $x_g$ bin we parametrised the LO $\kappa$ dependence 
and added it to the PS calculation.
A $\chi^2$ minimisation fit was performed to the data, yielding 
$\kappa = -0.011 \pm0.048{\rm (stat.)}^{+0.013}_{-0.003}{\rm (syst.)}$.
This corresponds to 95$\%$ c.l. limits of 
$-0.11 < \kappa < 0.08$ (preliminary). 

\section{Tests of Parity Violation at the \bbg Vertex}

New tests of parity-violation in strong interactions have recently been
proposed using polarized \ep \ra \qqg events~\cite{burrows}. 
The quark polar-angle distribution can be written:
$$
{{d\sigma} \over {d\cos\theta_b}} \propto 
(1-P_e A_e)(1+\alpha
\cos^2\theta_b)+2 A_P (P_e-A_e) \cos\theta_b 
$$
where $P_e$ is the electron polarisation, $A_e$ ($A_f$) is the
parity-violating electroweak coupling of the $Z^0$ to the initial (final)
state,  given by $A_i = 2v_ia_i/(v_i^2+a_i^2)$ in terms of
the vector $v_i$ and axial-vector $a_i$  couplings, and $A_P$ characterizes
the degree of parity violation. One can write $A_P$ = 
$A_f \cdot A^{QCD}_\theta$, where the second factor modulates the 
electroweak parity violation and can be calculated in QCD. Similarly, 
for the azimuthal angle $\chi$:
$$
{{d\sigma}\over {d\chi}} \propto (1-P_e A_e)(1+\beta
\cos 2\chi) - {\pi\over 2} A_P'  (P_e-A_e) \cos\chi 
$$
and $A_P'$ = $A_f \cdot A^{QCD}_\chi$. 
Given the SM value $A_b$ $\simeq$ 0.935, 
measurement of $A_P$ and $A_P'$ in 
\z0 \ra \bbg events allows one to test the 
QCD prediction for  $A^{QCD}_\theta$ and $A^{QCD}_\chi$.

Fig. 2 shows the observed $\cos\theta_b$ distributions.
The shaded histograms
show the estimated backgrounds, evaluated using JETSET, 
which are mostly $c\bar{c}g$ events.
A maximum likelihood fit yields 
$A^{QCD}_\theta$=$0.906\pm0.052\pm0.064$ (prelim.), 
consistent with the \oalpsq expectation of
0.93, evaluated using JETSET.
The $\chi$ distribution is shown in Fig.~3.
A corresponding fit yields
$A^{QCD}_\chi=-0.014 \pm0.035\pm0.002$ (prelim.), to be compared with the 
\oalpsq expectation of $-0.064$.
The asymmetry parameters are consistent with the expected degree of
parity violation in the SM, and we see no evidence for any anomalous effects.

\begin{figure}
\vspace*{-0.8cm}
\begin{center}
\epsfxsize=12cm
\epsfysize=8cm
\epsffile{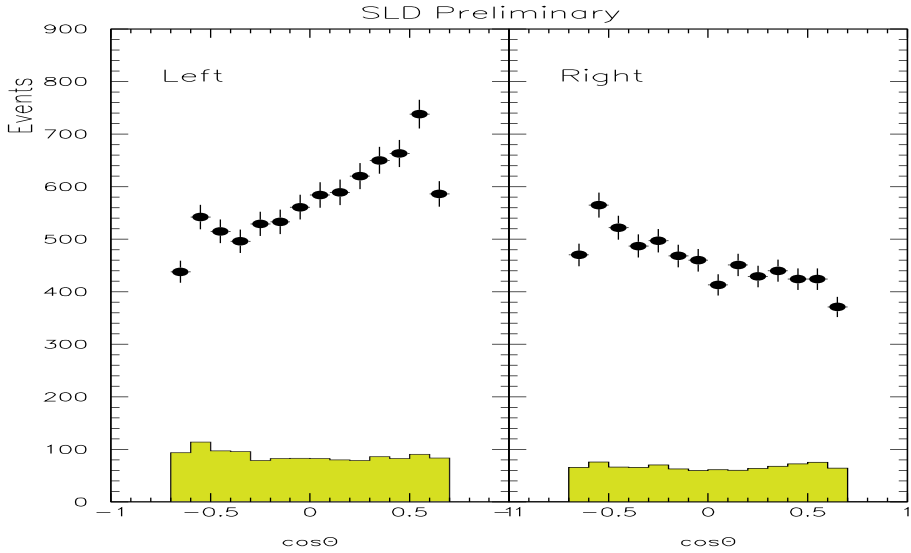}
\vspace*{-0.5cm}
\caption{
Distributions of cos$\theta_b$ (dots) for (a) left- and (b) 
right-handed polarised electrons.
}
\end{center}
\label{fig:theta}
\vspace*{-0.7cm}
\end{figure}

\begin{figure}
\begin{center}
\epsfxsize=12cm
\epsfysize=8cm
\epsffile{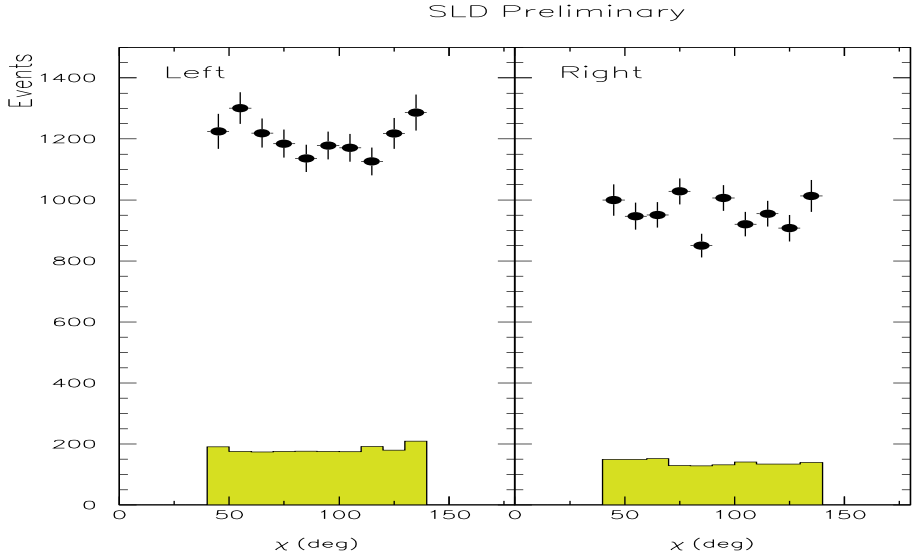}
\vspace*{-0.5cm}
\caption{
Distributions of $\chi$ (dots) for (a) left- and (b) 
right-handed polarised electrons.
}
\end{center}
\label{fig:chi}
\vspace*{-1cm}
\end{figure}

\section{T$_N$-odd Final-State Interactions}

Consider the polar angle, $\omega$, of the normal to the \bbg plane.  In PQCD 
one expects:
$$
{d\sigma\over{d\cos\omega}}\propto(1-P_e A_e)(1+\gamma \cos^2\omega)+
2A_T(P_e-A_e)\cos\omega
$$
Taking the left-right forward-backward asymmetry projects out the cos$\omega$ term.
This term is proportional to 
the triple product $\vec{\sigma}_Z \cdot (\vec{p_{b_i}} \times \vec{p_{b_j}})$, where
$\vec{\sigma}_Z$ is the $Z^0$ polarization vector.
When the vector product is ordered by jet momentum the term is $T_N$-odd and CP-even
(``$A_T^+$'').
Since the true time-reversed experiment is not performed non-zero contributions 
can arise from final-state interactions in the SM. 
A 1-loop SM calculation ~\cite{lance}
shows that $A_T^+$ is largest for $b\bar{b}g$
events, but is only $\sim$10$^{-5}$.
We have previously set limits on $A_T^+$ using events of all flavours~\cite{sldtodd}.
When the vector product is ordered by flavour, \ie $\vec{p_b} \times \vec{p_{\bar{b}}}$
the cos$\omega$ term is both $T_N$-odd and CP-odd (``$A_T^-$'').

For tagged \bbg events our measured left-right forward-backward asymmetries in the 
CP-even and odd cases are shown in Fig.~4.
They are both consistent with zero and we set 95\% c.l. limits on 
$T_N$-odd
asymmetries of $-0.038<A^+_T<0.014$ and $-0.077<A^-_T<0.011$, respectively (prelim.).

\begin{figure}
\vspace*{-0.8cm}
\begin{center}
\epsfxsize=12cm
\epsfysize=8cm
\epsffile{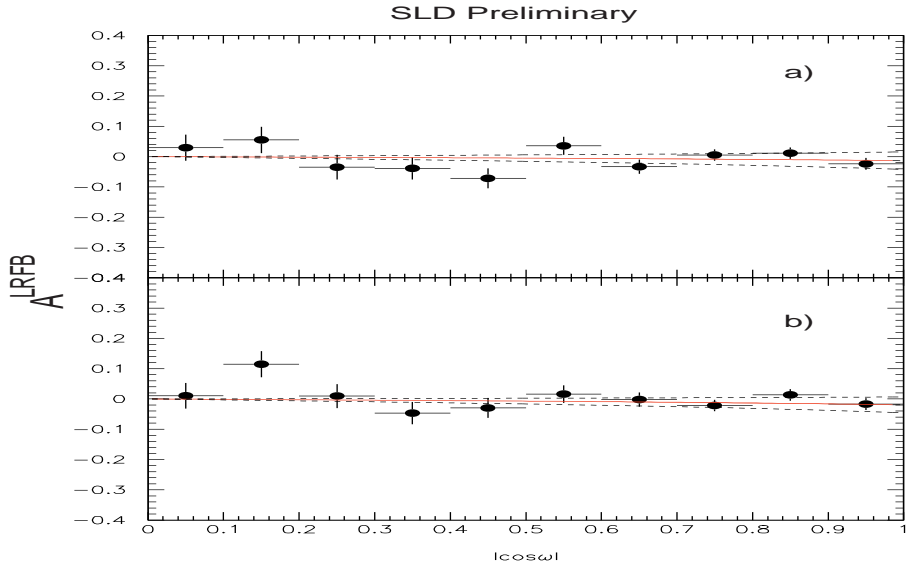}
\vspace*{-0.5cm}
\caption{
Left-right-forward-backward asymmetry vs. 
$|\cos\omega|$ for (a) CP-even, (b) CP-odd cases.
In each case the solid curve is the best fit, and the dashed curves 
correspond to the 95\% c.l. limits.
} 
\end{center}
\label{fig:todd}
\vspace*{-1cm}
\end{figure}

\vfill
\eject

\noindent
{\bf $*$ List of Authors}

%
% author list for inclusion in LaTeX documents
% using \author{} and \address{} commands
%
% Institution number definitions:
%
\begin{center}
\def\iADEL{$^{(1)}$}
\def\iAOMORI{$^{(2)}$}
\def\iBOLO{$^{(3)}$}
\def\iBRI{$^{(4)}$}
\def\iBRUN{$^{(5)}$}
\def\iBU{$^{(6)}$}
\def\iCINC{$^{(7)}$}
\def\iCOLO{$^{(8)}$}
\def\iCOLU{$^{(9)}$}
\def\iCSU{$^{(10)}$}
\def\iFERR{$^{(11)}$}
\def\iFRAS{$^{(12)}$}
\def\iILLI{$^{(13)}$}
\def\iJHU{$^{(14)}$}
\def\iLBL{$^{(15)}$}
\def\iLTU{$^{(16)}$}
\def\iMASS{$^{(17)}$}
\def\iMISSI{$^{(18)}$}
\def\iMIT{$^{(19)}$}
\def\iMOSCOW{$^{(20)}$}
\def\iNAGO{$^{(21)}$}
\def\iOREG{$^{(22)}$}
\def\iOXF{$^{(23)}$}
\def\iPADO{$^{(24)}$}
\def\iPERU{$^{(25)}$}
\def\iPISA{$^{(26)}$}
\def\iRAL{$^{(27)}$}
\def\iRUTG{$^{(28)}$}
\def\iSLAC{$^{(29)}$}
\def\iSOGA{$^{(30)}$}
\def\iSOONG{$^{(31)}$}
\def\iTENN{$^{(32)}$}
\def\iTOHO{$^{(33)}$}
\def\iUCSB{$^{(34)}$}
\def\iUCSC{$^{(35)}$}
\def\iUVIC{$^{(36)}$}
\def\iVAND{$^{(37)}$}
\def\iWASH{$^{(38)}$}
\def\iWISC{$^{(39)}$}
\def\iYALE{$^{(40)}$}

  \baselineskip=.75\baselineskip  
\mbox{Kenji  Abe\unskip,\iNAGO}
\mbox{Koya Abe\unskip,\iTOHO}
\mbox{T. Abe\unskip,\iSLAC}
\mbox{I. Adam\unskip,\iSLAC}
\mbox{T.  Akagi\unskip,\iSLAC}
\mbox{H. Akimoto\unskip,\iSLAC}
\mbox{N.J. Allen\unskip,\iBRUN}
\mbox{W.W. Ash\unskip,\iSLAC}
\mbox{D. Aston\unskip,\iSLAC}
\mbox{K.G. Baird\unskip,\iMASS}
\mbox{C. Baltay\unskip,\iYALE}
\mbox{H.R. Band\unskip,\iWISC}
\mbox{M.B. Barakat\unskip,\iLTU}
\mbox{O. Bardon\unskip,\iMIT}
\mbox{T.L. Barklow\unskip,\iSLAC}
\mbox{G.L. Bashindzhagyan\unskip,\iMOSCOW}
\mbox{J.M. Bauer\unskip,\iMISSI}
\mbox{G. Bellodi\unskip,\iOXF}
\mbox{A.C. Benvenuti\unskip,\iBOLO}
\mbox{G.M. Bilei\unskip,\iPERU}
\mbox{D. Bisello\unskip,\iPADO}
\mbox{G. Blaylock\unskip,\iMASS}
\mbox{J.R. Bogart\unskip,\iSLAC}
\mbox{G.R. Bower\unskip,\iSLAC}
\mbox{J.E. Brau\unskip,\iOREG}
\mbox{M. Breidenbach\unskip,\iSLAC}
\mbox{W.M. Bugg\unskip,\iTENN}
\mbox{D. Burke\unskip,\iSLAC}
\mbox{T.H. Burnett\unskip,\iWASH}
\mbox{P.N. Burrows\unskip,\iOXF}
\mbox{R.M. Byrne\unskip,\iMIT}
\mbox{A. Calcaterra\unskip,\iFRAS}
\mbox{D. Calloway\unskip,\iSLAC}
\mbox{B. Camanzi\unskip,\iFERR}
\mbox{M. Carpinelli\unskip,\iPISA}
\mbox{R. Cassell\unskip,\iSLAC}
\mbox{R. Castaldi\unskip,\iPISA}
\mbox{A. Castro\unskip,\iPADO}
\mbox{M. Cavalli-Sforza\unskip,\iUCSC}
\mbox{A. Chou\unskip,\iSLAC}
\mbox{E. Church\unskip,\iWASH}
\mbox{H.O. Cohn\unskip,\iTENN}
\mbox{J.A. Coller\unskip,\iBU}
\mbox{M.R. Convery\unskip,\iSLAC}
\mbox{V. Cook\unskip,\iWASH}
\mbox{R.F. Cowan\unskip,\iMIT}
\mbox{D.G. Coyne\unskip,\iUCSC}
\mbox{G. Crawford\unskip,\iSLAC}
\mbox{C.J.S. Damerell\unskip,\iRAL}
\mbox{M.N. Danielson\unskip,\iCOLO}
\mbox{M. Daoudi\unskip,\iSLAC}
\mbox{N. de Groot\unskip,\iBRI}
\mbox{R. Dell'Orso\unskip,\iPERU}
\mbox{P.J. Dervan\unskip,\iBRUN}
\mbox{R. de Sangro\unskip,\iFRAS}
\mbox{M. Dima\unskip,\iCSU}
\mbox{D.N. Dong\unskip,\iMIT}
\mbox{M. Doser\unskip,\iSLAC}
\mbox{R. Dubois\unskip,\iSLAC}
\mbox{B.I. Eisenstein\unskip,\iILLI}
\mbox{I.Erofeeva\unskip,\iMOSCOW}
\mbox{V. Eschenburg\unskip,\iMISSI}
\mbox{E. Etzion\unskip,\iWISC}
\mbox{S. Fahey\unskip,\iCOLO}
\mbox{D. Falciai\unskip,\iFRAS}
\mbox{C. Fan\unskip,\iCOLO}
\mbox{J.P. Fernandez\unskip,\iUCSC}
\mbox{M.J. Fero\unskip,\iMIT}
\mbox{K. Flood\unskip,\iMASS}
\mbox{R. Frey\unskip,\iOREG}
\mbox{J. Gifford\unskip,\iUVIC}
\mbox{T. Gillman\unskip,\iRAL}
\mbox{G. Gladding\unskip,\iILLI}
\mbox{S. Gonzalez\unskip,\iMIT}
\mbox{E.R. Goodman\unskip,\iCOLO}
\mbox{E.L. Hart\unskip,\iTENN}
\mbox{J.L. Harton\unskip,\iCSU}
\mbox{K. Hasuko\unskip,\iTOHO}
\mbox{S.J. Hedges\unskip,\iBU}
\mbox{S.S. Hertzbach\unskip,\iMASS}
\mbox{M.D. Hildreth\unskip,\iSLAC}
\mbox{J. Huber\unskip,\iOREG}
\mbox{M.E. Huffer\unskip,\iSLAC}
\mbox{E.W. Hughes\unskip,\iSLAC}
\mbox{X. Huynh\unskip,\iSLAC}
\mbox{H. Hwang\unskip,\iOREG}
\mbox{M. Iwasaki\unskip,\iOREG}
\mbox{D.J. Jackson\unskip,\iRAL}
\mbox{P. Jacques\unskip,\iRUTG}
\mbox{J.A. Jaros\unskip,\iSLAC}
\mbox{Z.Y. Jiang\unskip,\iSLAC}
\mbox{A.S. Johnson\unskip,\iSLAC}
\mbox{J.R. Johnson\unskip,\iWISC}
\mbox{R.A. Johnson\unskip,\iCINC}
\mbox{T. Junk\unskip,\iSLAC}
\mbox{R. Kajikawa\unskip,\iNAGO}
\mbox{M. Kalelkar\unskip,\iRUTG}
\mbox{Y. Kamyshkov\unskip,\iTENN}
\mbox{H.J. Kang\unskip,\iRUTG}
\mbox{I. Karliner\unskip,\iILLI}
\mbox{H. Kawahara\unskip,\iSLAC}
\mbox{Y.D. Kim\unskip,\iSOGA}
\mbox{M.E. King\unskip,\iSLAC}
\mbox{R. King\unskip,\iSLAC}
\mbox{R.R. Kofler\unskip,\iMASS}
\mbox{N.M. Krishna\unskip,\iCOLO}
\mbox{R.S. Kroeger\unskip,\iMISSI}
\mbox{M. Langston\unskip,\iOREG}
\mbox{A. Lath\unskip,\iMIT}
\mbox{D.W.G. Leith\unskip,\iSLAC}
\mbox{V. Lia\unskip,\iMIT}
\mbox{C.Lin\unskip,\iMASS}
\mbox{M.X. Liu\unskip,\iYALE}
\mbox{X. Liu\unskip,\iUCSC}
\mbox{M. Loreti\unskip,\iPADO}
\mbox{A. Lu\unskip,\iUCSB}
\mbox{H.L. Lynch\unskip,\iSLAC}
\mbox{J. Ma\unskip,\iWASH}
\mbox{M. Mahjouri\unskip,\iMIT}
\mbox{G. Mancinelli\unskip,\iRUTG}
\mbox{S. Manly\unskip,\iYALE}
\mbox{G. Mantovani\unskip,\iPERU}
\mbox{T.W. Markiewicz\unskip,\iSLAC}
\mbox{T. Maruyama\unskip,\iSLAC}
\mbox{H. Masuda\unskip,\iSLAC}
\mbox{E. Mazzucato\unskip,\iFERR}
\mbox{A.K. McKemey\unskip,\iBRUN}
\mbox{B.T. Meadows\unskip,\iCINC}
\mbox{G. Menegatti\unskip,\iFERR}
\mbox{R. Messner\unskip,\iSLAC}
\mbox{P.M. Mockett\unskip,\iWASH}
\mbox{K.C. Moffeit\unskip,\iSLAC}
\mbox{T.B. Moore\unskip,\iYALE}
\mbox{M.Morii\unskip,\iSLAC}
\mbox{D. Muller\unskip,\iSLAC}
\mbox{V. Murzin\unskip,\iMOSCOW}
\mbox{T. Nagamine\unskip,\iTOHO}
\mbox{S. Narita\unskip,\iTOHO}
\mbox{U. Nauenberg\unskip,\iCOLO}
\mbox{H. Neal\unskip,\iSLAC}
\mbox{M. Nussbaum\unskip,\iCINC}
\mbox{N. Oishi\unskip,\iNAGO}
\mbox{D. Onoprienko\unskip,\iTENN}
\mbox{L.S. Osborne\unskip,\iMIT}
\mbox{R.S. Panvini\unskip,\iVAND}
\mbox{C.H. Park\unskip,\iSOONG}
\mbox{T.J. Pavel\unskip,\iSLAC}
\mbox{I. Peruzzi\unskip,\iFRAS}
\mbox{M. Piccolo\unskip,\iFRAS}
\mbox{L. Piemontese\unskip,\iFERR}
\mbox{K.T. Pitts\unskip,\iOREG}
\mbox{R.J. Plano\unskip,\iRUTG}
\mbox{R. Prepost\unskip,\iWISC}
\mbox{C.Y. Prescott\unskip,\iSLAC}
\mbox{G.D. Punkar\unskip,\iSLAC}
\mbox{J. Quigley\unskip,\iMIT}
\mbox{B.N. Ratcliff\unskip,\iSLAC}
\mbox{T.W. Reeves\unskip,\iVAND}
\mbox{J. Reidy\unskip,\iMISSI}
\mbox{P.L. Reinertsen\unskip,\iUCSC}
\mbox{P.E. Rensing\unskip,\iSLAC}
\mbox{L.S. Rochester\unskip,\iSLAC}
\mbox{P.C. Rowson\unskip,\iCOLU}
\mbox{J.J. Russell\unskip,\iSLAC}
\mbox{O.H. Saxton\unskip,\iSLAC}
\mbox{T. Schalk\unskip,\iUCSC}
\mbox{R.H. Schindler\unskip,\iSLAC}
\mbox{B.A. Schumm\unskip,\iUCSC}
\mbox{J. Schwiening\unskip,\iSLAC}
\mbox{S. Sen\unskip,\iYALE}
\mbox{V.V. Serbo\unskip,\iSLAC}
\mbox{M.H. Shaevitz\unskip,\iCOLU}
\mbox{J.T. Shank\unskip,\iBU}
\mbox{G. Shapiro\unskip,\iLBL}
\mbox{D.J. Sherden\unskip,\iSLAC}
\mbox{K.D. Shmakov\unskip,\iTENN}
\mbox{C. Simopoulos\unskip,\iSLAC}
\mbox{N.B. Sinev\unskip,\iOREG}
\mbox{S.R. Smith\unskip,\iSLAC}
\mbox{M.B. Smy\unskip,\iCSU}
\mbox{J.A. Snyder\unskip,\iYALE}
\mbox{H. Staengle\unskip,\iCSU}
\mbox{A. Stahl\unskip,\iSLAC}
\mbox{P. Stamer\unskip,\iRUTG}
\mbox{H. Steiner\unskip,\iLBL}
\mbox{R. Steiner\unskip,\iADEL}
\mbox{M.G. Strauss\unskip,\iMASS}
\mbox{D. Su\unskip,\iSLAC}
\mbox{F. Suekane\unskip,\iTOHO}
\mbox{A. Sugiyama\unskip,\iNAGO}
\mbox{S. Suzuki\unskip,\iNAGO}
\mbox{M. Swartz\unskip,\iJHU}
\mbox{A. Szumilo\unskip,\iWASH}
\mbox{T. Takahashi\unskip,\iSLAC}
\mbox{F.E. Taylor\unskip,\iMIT}
\mbox{J. Thom\unskip,\iSLAC}
\mbox{E. Torrence\unskip,\iMIT}
\mbox{N.K. Toumbas\unskip,\iSLAC}
\mbox{T. Usher\unskip,\iSLAC}
\mbox{C. Vannini\unskip,\iPISA}
\mbox{J. Va'vra\unskip,\iSLAC}
\mbox{E. Vella\unskip,\iSLAC}
\mbox{J.P. Venuti\unskip,\iVAND}
\mbox{R. Verdier\unskip,\iMIT}
\mbox{P.G. Verdini\unskip,\iPISA}
\mbox{D.L. Wagner\unskip,\iCOLO}
\mbox{S.R. Wagner\unskip,\iSLAC}
\mbox{A.P. Waite\unskip,\iSLAC}
\mbox{S. Walston\unskip,\iOREG}
\mbox{S.J. Watts\unskip,\iBRUN}
\mbox{A.W. Weidemann\unskip,\iTENN}
\mbox{E. R. Weiss\unskip,\iWASH}
\mbox{J.S. Whitaker\unskip,\iBU}
\mbox{S.L. White\unskip,\iTENN}
\mbox{F.J. Wickens\unskip,\iRAL}
\mbox{B. Williams\unskip,\iCOLO}
\mbox{D.C. Williams\unskip,\iMIT}
\mbox{S.H. Williams\unskip,\iSLAC}
\mbox{S. Willocq\unskip,\iMASS}
\mbox{R.J. Wilson\unskip,\iCSU}
\mbox{W.J. Wisniewski\unskip,\iSLAC}
\mbox{J. L. Wittlin\unskip,\iMASS}
\mbox{M. Woods\unskip,\iSLAC}
\mbox{G.B. Word\unskip,\iVAND}
\mbox{T.R. Wright\unskip,\iWISC}
\mbox{J. Wyss\unskip,\iPADO}
\mbox{R.K. Yamamoto\unskip,\iMIT}
\mbox{J.M. Yamartino\unskip,\iMIT}
\mbox{X. Yang\unskip,\iOREG}
\mbox{J. Yashima\unskip,\iTOHO}
\mbox{S.J. Yellin\unskip,\iUCSB}
\mbox{C.C. Young\unskip,\iSLAC}
\mbox{H. Yuta\unskip,\iAOMORI}
\mbox{G. Zapalac\unskip,\iWISC}
\mbox{R.W. Zdarko\unskip,\iSLAC}
\mbox{J. Zhou\unskip.\iOREG}

\it
  \vskip \baselineskip                   % \bigskip did not work
  \centerline{(The SLD Collaboration)}   % include collaboration name
  \vskip \baselineskip        
  \baselineskip=.75\baselineskip   % shrink the interline spacing
\iADEL
  Adelphi University, Garden City, New York 11530, \break
\iAOMORI
  Aomori University, Aomori , 030 Japan, \break
\iBOLO
  INFN Sezione di Bologna, I-40126, Bologna, Italy, \break
\iBRI
  University of Bristol, Bristol, U.K., \break
\iBRUN
  Brunel University, Uxbridge, Middlesex, UB8 3PH United Kingdom, \break
\iBU
  Boston University, Boston, Massachusetts 02215, \break
\iCINC
  University of Cincinnati, Cincinnati, Ohio 45221, \break
\iCOLO
  University of Colorado, Boulder, Colorado 80309, \break
\iCOLU
  Columbia University, New York, New York 10533, \break
\iCSU
  Colorado State University, Ft. Collins, Colorado 80523, \break
\iFERR
  INFN Sezione di Ferrara and Universita di Ferrara, I-44100 Ferrara, Italy, \break
\iFRAS
  INFN Lab. Nazionali di Frascati, I-00044 Frascati, Italy, \break
\iILLI
  University of Illinois, Urbana, Illinois 61801, \break
\iJHU
  Johns Hopkins University,  Baltimore, Maryland 21218-2686, \break
\iLBL
  Lawrence Berkeley Laboratory, University of California, Berkeley, California 94720, \break
\iLTU
  Louisiana Technical University, Ruston,Louisiana 71272, \break
\iMASS
  University of Massachusetts, Amherst, Massachusetts 01003, \break
\iMISSI
  University of Mississippi, University, Mississippi 38677, \break
\iMIT
  Massachusetts Institute of Technology, Cambridge, Massachusetts 02139, \break
\iMOSCOW
  Institute of Nuclear Physics, Moscow State University, 119899, Moscow Russia, \break
\iNAGO
  Nagoya University, Chikusa-ku, Nagoya, 464 Japan, \break
\iOREG
  University of Oregon, Eugene, Oregon 97403, \break
\iOXF
  Oxford University, Oxford, OX1 3RH, United Kingdom, \break
\iPADO
  INFN Sezione di Padova and Universita di Padova I-35100, Padova, Italy, \break
\iPERU
  INFN Sezione di Perugia and Universita di Perugia, I-06100 Perugia, Italy, \break
\iPISA
  INFN Sezione di Pisa and Universita di Pisa, I-56010 Pisa, Italy, \break
\iRAL
  Rutherford Appleton Laboratory, Chilton, Didcot, Oxon OX11 0QX United Kingdom, \break
\iRUTG
  Rutgers University, Piscataway, New Jersey 08855, \break
\iSLAC
  Stanford Linear Accelerator Center, Stanford University, Stanford, California 94309, \break
\iSOGA
  Sogang University, Seoul, Korea, \break
\iSOONG
  Soongsil University, Seoul, Korea 156-743, \break
\iTENN
  University of Tennessee, Knoxville, Tennessee 37996, \break
\iTOHO
  Tohoku University, Sendai 980, Japan, \break
\iUCSB
  University of California at Santa Barbara, Santa Barbara, California 93106, \break
\iUCSC
  University of California at Santa Cruz, Santa Cruz, California 95064, \break
\iUVIC
  University of Victoria, Victoria, British Columbia, Canada V8W 3P6, \break
\iVAND
  Vanderbilt University, Nashville,Tennessee 37235, \break
\iWASH
  University of Washington, Seattle, Washington 98105, \break
\iWISC
  University of Wisconsin, Madison,Wisconsin 53706, \break
\iYALE
  Yale University, New Haven, Connecticut 06511. \break

\rm
%
%  }   % end of address list

\end{center}

\end{document}